# Efficient Rewirings for Enhancing Synchronizability of Dynamical Networks

Ali Ajdari Rad, Mahdi Jalili, and Martin Hasler

Ecole Polytechnique Fédérale de Lausanne, Laboratory of Nonlinear Systems

School of Computer & Communication Science, CH1015 Lausanne, Switzerland

E-mail addresses: {ali.ajdarirad, mahdi.jalili, martin.hasler}@epfl.ch

## Abstract

In this paper, we present an algorithm for optimizing synchronizability of complex dynamical networks. Starting with an undirected and unweighted network, we end up with an undirected and unweighted network with the same number of nodes and edges having enhanced synchronizability. To this end, based on some network properties, rewirings, i.e. eliminating an edge and creating a new edge elsewhere, are performed iteratively avoiding always self-loops and multiple edges between the same nodes. We show that the method is able to enhance the synchronizability of networks of any size and topological properties in a small number of steps that scales with the network size. For numerical simulations, an optimization algorithm based on simulated annealing is used. Also, the evolution of different topological properties of the network such as distribution of node degree, node and edge betweenness centrality is tracked with the iteration steps. We use networks such as scale-free, Strogatz-Watts and random to start with and we show that regardless of the initial network, the final optimized network becomes homogenous. In other words, in the network with high synchronizability, parameters such as degree, shortest distance, node and edge betweenness centrality are almost homogeneously distributed. Also, parameters such as maximum node and edge betweenness centrality are small for the rewired network. Although we take the eigenratio of the Laplacian as the target function for optimization, we will show that it is also possible to choose other appropriate target functions exhibiting almost the same performance. Furthermore, we show that even if the network is optimized taking into account another interpretation of synchronizability, i.e. synchronization cost, the optimal network will have the same synchronization properties. Indeed, in networks with optimized synchronizability, different interpretations of synchronizability coincide. The optimized networks are Ramanujan graphs, and thus, this rewiring algorithm could be used to produce Ramanujan graphs of any size and average degree.

## Lead paragraph

In recent years, the concept of collective behavior has been recognized in many branches of science and subsequently many studies have been initiated to investigate various types of collective behavior. Synchronization, the most striking form of collective behavior, has been in the center of interest of these studies [1]. Synchronization is not defined uniquely and a particular definition is adopted for each type of application [2]. Complete (identical) synchronization is the strongest type of synchronization, which is achieved when the coupling between dynamical systems is sufficiently strong. Some works have tried to formulize the problem of complete synchronization and relate it to some properties of the connection graph and the dynamics of the individual dynamical systems [3,4]. Synchronizability of a dynamical network is the ease by which the individual dynamical systems, sitting on the nodes of the network and interacting through the edges, can be synchronized. There are various possible interpretations of synchronizability [5]. For example, one may argue that network $N_1$ is more synchronizable than network $N_2$, if for a larger range of parameters, it is possible to synchronize $N_1$ compared to $N_2$. Another interpretation could be that $N_1$ is more synchronizable than $N_2$, if less effort has to be made to achieve synchronization in $N_1$ than in $N_2$. Networks with a particular topological property might have better synchronizability than networks with another topological property. Designing networks with high synchronizability has various potential applications. Technological networks with desirable synchronizability are typical examples where properly assigned interaction between dynamical units is of high importance. In particular, designing interaction schemes to optimize the performance of computational tasks based on the synchronization of processes in computer networks [6] and designing networks with optimized synchronizability for sensor networks where synchronization is used as a mechanism for consensus [7]. Neuronal networks are other prototypic examples, where studying optimal synchronizability may advance our understanding of their organizing principles. Other examples could be biological and social networks. In this paper, we will introduce a rewiring algorithm to enhance the synchronizability of dynamical networks. The procedure will be supported by various numerical simulations.

## Introduction

Complex networks such as Internet, World Wide Web, engineering, social, biological and economical networks have been extensively studied in recent years and the publication volume is growing at a high rate [8-11]. Research has shown that many real-world networks from physics to biology, engineering and sociology have some common structural properties [12]. Watts and Strogatz in their seminal work [13] observed that many real-world networks exhibit, in general, two properties; namely, they show the Strogatz-Watts property meaning that the average shortest path length scales logarithmically with the network size, a property of random networks. Furthermore, many real-world networks have high clustering (or transitivity); meaning that there is an increased probability that two nodes will be connected directly to one another if they have another neighboring node in common. Many real-world networks have also a power-law node-degree distribution, meaning that when a new node is added to the networks, the probability of being connected to other nodes is proportional to their degrees; the higher the degree of the node the higher the probability of being connected. Barabási and Albert reported this fact in their work and proposed a preferential attachment algorithm to construct such networks [14].

An interesting phenomenon that is observed in complex networks is collective behavior, *synchronization* [1, 15]. Due to strong enough interaction between dynamical systems, they can adjust their motion, i.e. they can synchronize. Synchronization is often encountered in living systems such as circadian rhythm, phase locking respiration with mechanical ventilator, phase locking of chicken embrion heart cells with external stimuli, interaction of the sinus node with ectopic pacemakers, synchronization of oscillations of human insulin secretion and glucose infusion, locking of spiking from electroreceptors of a paddlefish to weak external electromagnetic field, synchronization of heart rate by external audio or visual stimuli and synchronization of neurons in the brain [1]. Indeed, the tendency to achieve common rhythms of mutual behavior, or in other words, the tendency to synchronization, is an important feature in our living world. During the last couple of decades the notion of synchronization has been generalized to the case of interacting chaotic oscillators [16], which has led to different concepts of synchronization such as complete (identical), phase and generalized synchronization [17]. Synchronization is possible if at least two dynamical systems interact but it also happens in ensembles including hundreds, thousands, millions and even more individual dynamical units. The early works on synchronization of dynamical systems were concerned with only a small number of coupled oscillators, but many real-world systems where synchronization is relevant, consist of a large number of dynamical systems interacting with a complex coupling structure.

One important issue in studying the synchronization of dynamical networks is their synchronizability. For many applications, it is desired to have networks with high synchronizability. For examples, sensor networks with high degrees of synchronizability have faster convergence time than networks with lower synchronizability [18]. In general, there are two methods for enhancing synchronizability of dynamical networks: assigning proper connection weights [5, 19-23] and rewiring the links [24, 25]. In this work, we design an efficient algorithm for performing rewirings to obtain a network with enhanced synchronizability. Starting with an unweighted and undirected network and by considering some properties of the network, at each iteration step a proper rewiring is performed. The proposed optimization algorithm is fast and in - relatively few steps it is able to find a network structure with high synchronizability. Numerical simulations are performed to support the algorithm.

## Synchronizability of dynamical networks

Let us consider an undirected and unweighted network with $N$ nodes with the dynamics of motion as

$$\dot{x}_i = F(x_i) - \sigma \sum_{j=1}^{N} L_{ij} H x_j \quad ; \quad i = 1, 2, ..., N. \tag{1}$$

where $x_i \in \mathbb{R}^d$ is a $d$–dimensional vector of state variables of the $i$-th individual dynamical system and $\sigma$ is the (constant) diffusive coupling strength. $F : \mathbb{R}^d \to \mathbb{R}^d$ defines the individual system's state equation. $L = (L_{ij})$ that is called *Laplacian* is a symmetric matrix with vanishing row-sums and positive diagonal entries, i.e. $L_{ij} = L_{ji}$ for all pairs of $(i,j)$, $L_{ij} < 0$ for $i \neq j$, and $\sum_{j=1}^{N} L_{ij} = 0$ for all $i$. The nonzero elements of the $d \times d$ matrix $H$ determine the coupling between the various states of individual dynamical systems that are interacting.

The local stability of the synchronization manifold $x_1 = x_2 = \ldots = x_N$ can be studied in the formalism of master-stability-function [4]. The variational equations of the dynamical network (1) along a synchronized solution $x_1(t) = x_2(t) = \ldots = x_N(t) = s(t)$ can be written as

$$\dot{\zeta}_i = DF(s)\zeta_i - \sigma \sum_{\substack{j=1 \\ j \neq i}}^{N} L_{ij} H \zeta_j \quad ; \quad i = 1, 2, ..., N, \tag{2}$$

where $D$ stands for the Jacobian operator. One can write the symmetric matrix $L$ as $L = \Gamma \Omega \Gamma^T$, where $\Omega$ is a diagonal matrix of real eigenvalues of $L$ and $\Gamma$ is the orthogonal matrix whose columns are the corresponding real eigenvectors of $L$. Let us define $\zeta = (\zeta_1, \zeta_2, \ldots, \zeta_N) = \eta \Gamma^T$, where $\eta = (\eta_1, \eta_2, \ldots, \eta_N)$. Then, Eq. (2) is equivalent to

$$\dot{\boldsymbol{\eta}}_i = DF(s)\boldsymbol{\eta}_i - \sigma\lambda_i H\boldsymbol{\eta}_i \quad ; \quad i = 1, 2, ..., N, \qquad (3)$$

where $\lambda_i$, $i = 1,...N$, are the eigenvalues of $L$, ordered as $0 = \lambda_1 \leq \lambda_2 \leq ... \leq \lambda_N$, in which $\lambda_1 = 0$ is associated with the synchronization manifold. The largest Lyapunov exponent of Eq. (3), $\Lambda(\sigma\lambda_i)$, called master-stability-function [4], yields a necessary condition for the local stability of the synchronization manifold. If the synchronization manifold is stable, we must have $\Lambda(\sigma\lambda_i) < 0$, $\forall$ $i \geq 2$. Synchronizability of a dynamical network is the ease by which synchronization can be achieved. There is no single interpretation of synchronizability and a particular choice is adopted for each study [5]. For a number of systems such as $x$–coupled Rössler systems, the master-stability-function is negative only within a bounded interval $(a_1,a_2)$ [4]. Requiring all coupling strengths to lie within such an interval, i.e. $a_1 < \sigma\lambda_2 \leq ... \leq \sigma\lambda_N < a_2$, one concludes that the synchronization manifold can only be locally stable if

$$\frac{\lambda_N}{\lambda_2} < \frac{a_2}{a_1}. \qquad (4)$$

As it can be seen in Eq. (4) the left-hand side of the inequality depends solely on the structure of the network, while the right-hand side depends on the dynamics of the individual systems and the coupling configuration. One of the interpretations of synchronizability points out that the larger the range of connection strength stabilizing the synchronization manifold, the better the synchronizability of the network [5]. Therefore, Eq. (4) relates the synchronizability to the eigenratio $\lambda_N/\lambda_2$, and concludes that the smaller the eigenratio $\lambda_N/\lambda_2$ of a network, the better its synchronizability. This interpretation of synchronizability has been extensively used in the literature [19, 22, 25-27], even though it is linked to the situation where the master-stability-function is negative only in a finite interval, which is by far not always the case. However, other measures of synchronizability often go hand in hand with $\lambda_N/\lambda_2$.

## Rewiring as a mechanism to optimize $\lambda_N/\lambda_2$

Our aim is to build networks with predetermined size and average degree having optimal synchronization properties, i.e. minimal $\lambda_N/\lambda_2$. Let us first give an overview over the synchronizability of some well-known networks. In general, random networks have better synchronizability than regular networks, which is mainly due to the shorter average distance. Strogatz-Watts networks have, in general, better synchronization properties than scale-free networks [28-31]. Although average distance is an important factor determining the synchronizability of dynamical networks it might happen that networks with higher average distance have better synchronization properties than those with shorter distance [27]. It has been shown that the synchronizability is enhanced by decreasing the heterogeneity in the distribution of betweenness centrality [32]. In general, heterogeneity of the network is one of the most influential factors determining its synchronizability, the less heterogeneous the network the better its synchronizability [26].

Considering an undirected and unweighted network with $N$ nodes and average degree $<k>$, we would like to obtain an undirected and unweighted network with the same number of nodes and average degree, and thus the same number of edges, and with enhanced synchronizability, i.e. minimized eigenratio $\lambda_N/\lambda_2$. Donetti *et al.* proposed a simulated annealing based optimization algorithm to minimize $\lambda_N/\lambda_2$ [24, 25, 33]. In their proposed algorithm, at each step, a number of rewiring trials is randomly extracted from an exponential distribution. Each of them consists in removing a randomly selected link, and introducing a new one joining two random nodes. Then, the attempted rewiring is (*i*) rejected if the updated network is disconnected, or has a self-loop or multiple edges between the same nodes, otherwise, (*ii*) accepted if eigenratio $\lambda_N/\lambda_2$ of new network is less than the previous one or, (*iii*) accepted according to a probability measure. The process is iterated until there is no change during some successive steps, assuming that a relatively good local minimum of $\lambda_N/\lambda_2$ has been found. Although this method is powerful in finding a network topology with high synchronizability it is very expensive to perform and the completely random rewiring strategy limits its application to relatively small networks.

Wang *et al.* proposed a method using a heuristic memory based on tabu search to maximize network resilience [34]. Simultaneously, the eigenratio $\lambda_N/\lambda_2$ of the network has also been studied. By iterative random rewirings and a prescribed stop condition, they have tried to optimize the network. Fallat & Kirkland have proposed a graph-theoretical approach to maximize $\lambda_2$ over the set of trees of fixed diameter [35].

Ghosh & Boyd [36] have proposed a convex optimization method for growing well-connected networks. They proposed a heuristic greedy perturbation algorithm for adding proper edges to a base network that has maximum effect on increasing $\lambda_2$. If $\boldsymbol{u_2}$ is the eigenvector with unit norm corresponding to $\lambda_2$, then $\boldsymbol{u_2}\boldsymbol{u_2}^T$ is a super-gradient of $\lambda_2$ at the Laplacian matrix $L$, i.e., for any symmetric matrix $Y$ we have [36]

$$\lambda_2(L+Y) \leq \lambda_2(L) + trace(Y\boldsymbol{u}_2\boldsymbol{u}_2^T). \tag{6}$$

If $\lambda_2$ is isolated, i.e., $\lambda_1 < \lambda_2 < \lambda_3$, then $\lambda_2(L)$ is an analytic function of $L$. In this case, the super-gradient is the gradient, i.e.,

$$\lambda_2(L+Y) - \lambda_2(L) = trace(Y_{e_{ij}}\boldsymbol{u}_2\boldsymbol{u}_2^T) = (\boldsymbol{u}_{2i} - \boldsymbol{u}_{2j})^2, \tag{7}$$

where $\boldsymbol{u}_2$ is the unique normalized eigenvector (up to a sign flip) corresponding to $\lambda_2$, $e_{ij}$ is the added edge and $Y_{e_{ij}}$ is the Laplacian matrix of a symmetric adjacency matrix with 1 in the elements corresponding to $e_{ij}$ and zero elsewhere. In other words, when $\lambda_2(L)$ is isolated, $(\boldsymbol{u}_{2i} - \boldsymbol{u}_{2j})^2$ gives the first-order approximation of the increase in $\lambda_2(L)$ if edge $e_{ij}$ is added to the network. Then, they concluded that adding a non-existing edge that maximizes $(\boldsymbol{u}_{2i} - \boldsymbol{u}_{2j})^2$ seems to be a good strategy to increase $\lambda_2$ effectively [36].

In spectral graph theory $\lambda_N$ is often related to the maximum degree of the graph, i.e. $\lambda_N \in [k_{max}, 2k_{max}]$. Anderson and Morley showed that $\lambda_N \leq \max\{k_i + k_j\}$ where the $i$-th and the $j$-th nodes are adjacent [37]. Intuitively, one might try to decrease $k_{max}$ or $\max\{k_i + k_j\}$ to decrease $\lambda_N$.

Here we propose a rewiring algorithm, which takes advantages of graph structural properties to decide which edges are to be disconnected and which are the new connections. Considering a network with $N$ nodes and average degree $<k>$, the algorithm consists of the following steps:

The eigenratio $\lambda_N/\lambda_2$ of the network is calculated and in the first iteration step $(\lambda_N/\lambda_2)_{min} = \lambda_N/\lambda_2$.

For each edge $e_{ij}$ (connecting the $i$-th and $j$-th nodes) of the network the quantity $E_{cut,ij} = (k_i + k_j)$ is calculated, where $k_i$ is the degree of the $i$-th node. $E_{cut}$ is used to choose one edge for disconnecting, i.e. the probability of choosing an edge for disconnection is proportional to $\exp(E_{cut})$.

For each pair of non-adjacent nodes $i$ and $j$, the quantity $E_{connect,ij} = (\boldsymbol{u}_{2i} - \boldsymbol{u}_{2j})^2$ is calculated, which is used for choosing a pair of non-adjacent nodes to connect an edge between them. The probability of creating an edge between the $i$-th and the $j$-th (non-adjacent) node is proportional to $\exp(E_{connect,ij})$.

After rewiring, the cost function $\lambda_N/\lambda_2$ of the new network is calculated. Then

(*i*). If the network is disconnected, the rewiring is rejected, otherwise,

(*ii*). if the eigenratio of the new network $(\lambda_N/\lambda_2)_{new}$ is less than the eigenratio of the old network $(\lambda_N/\lambda_2)_{old}$, the rewiring is accepted and $(\lambda_N/\lambda_2)_{min} = (\lambda_N/\lambda_2)_{new}$,

(*iii*). if $(\lambda_N/\lambda_2)_{new} > (\lambda_N/\lambda_2)_{old}$, the rewiring is accepted with the probability of $\min(1, \max(0, THR - ((\lambda_N/\lambda_2)_{new} - (\lambda_N/\lambda_2)_{min})))$. THR is a threshold variable which is initially set to zero and in each step when the rewiring is rejected, it is increased by $d_{THR}/\log(T+1)$, where $d_{THR}$ is a constant and $T$ is the number of iterations performed. When the rewiring is accepted, THR is reset to zero. This procedure, which indeed is an extension to the simulated annealing approach, helps the algorithm to avoid getting trapped by a local minimum.

The algorithm is stopped after a predetermined number of iterations.

Next we will support this procedure by numerical simulations on some sample networks.

## Simulation results

We give numerical simulations of optimization of the eigenratio $\lambda_N/\lambda_2$ for different classes of networks. To this end, scale-free and Strogatz-Watts networks are considered. Scale-free networks are constructed by the following algorithm [19]. Starting with a network of $m + 1$ all-to-all connected nodes, at each step a new node is added to the network and is connected to $m$ other nodes. Such an edge connects to old nodes $i$ with probability $P_i = (k_i + B)/\sum_j (k_j + B)$, where $k_i$ is the degree of node $i$ and $B$ is a tunable real parameter controlling the heterogeneity of the network [19]. Strogatz-Watts networks are constructed using Watts-Strogatz algorithm [13] as follows. Considering a ring network with $N$ nodes, each connected to its $m$–nearest neighbors by undirected edges, the edges of each node are randomly rewired to other nodes of the network with a probability $P$, avoiding duplications of edges. Note that for the case with $P = 1$, a fully random network is obtained.

The optimization algorithm was applied to scale-free and Strogatz-Watts networks with $N = 200$ and $<k> = 6$. For the optimization algorithm we fixed $d_{THR} = 0.5$ and the results were averaged over 10 realizations. Figure 1 shows the eigenratio $\lambda_N/\lambda_2$ as a function of iteration steps. It shows that the eigenratio is exponentially decreasing as the algorithm proceeds and a dramatic decrease is obtained by introducing only a few rewirings. Our experience showed that about $2N$ iteration steps is enough to reach an asymptotic behavior. It is worth mentioning that the optimized network is largely independent of the initial network and no matter what the initial network is, the final network with optimized synchronizability have the same properties. For networks of small size, where the optimal network topology is known, our algorithm always finds the optimal network. An example of an optimized network where the initial network was a Strogatz-Watts network with $N = 50$ and $<k> = 4$ is depicted in Figure 2.

We also performed a more systematic analysis on the contribution of $\lambda_N$ and $\lambda_2$ on the optimization process. Figure 3 shows the profile of $\lambda_2$ and $\lambda_N$ as a function of the eigenratio $\lambda_N/\lambda_2$ during the optimization process for different networks with $N = 200$ and $<k> = 6$. As it is seen, in scale-free networks, in general, the optimization process influences $\lambda_N$ more than $\lambda_2$. In contrary, in Strogatz-Watts and random networks maximizing $\lambda_2$ plays the main role in the optimization of the eigenratio $\lambda_N/\lambda_2$. Indeed, since $\lambda_N$ has large values for heterogeneous networks [27], such as scale-free networks with small values of $B$, the optimization algorithm should first put much more effort to change the topology in such a manner as to reduce $\lambda_N$ than to try to maximize $\lambda_2$.

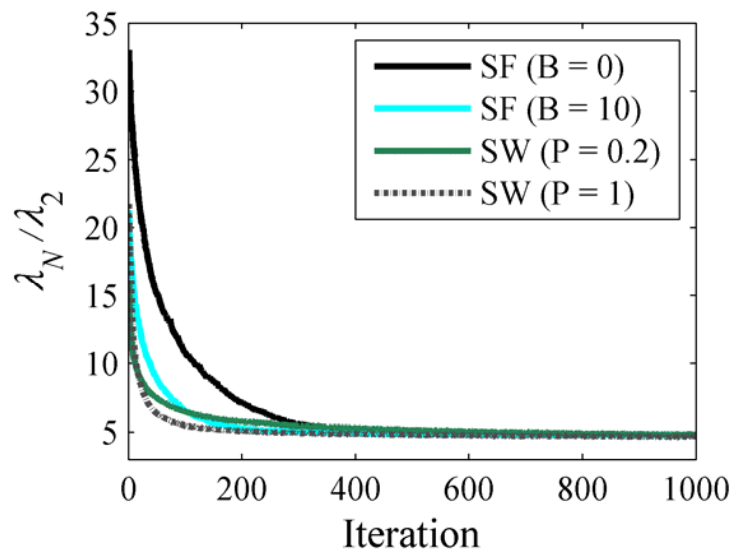

**Figure 1. The eigenratio $\lambda_N/\lambda_2$ as a function of iteration steps for scale-free (SF) and Strogatz-Watts (SW) network with $N = 200$ and $<k> = 6$. The optimization target is to minimize the eigenratio $\lambda_N/\lambda_2$. Data is averaged over 10 realizations.**

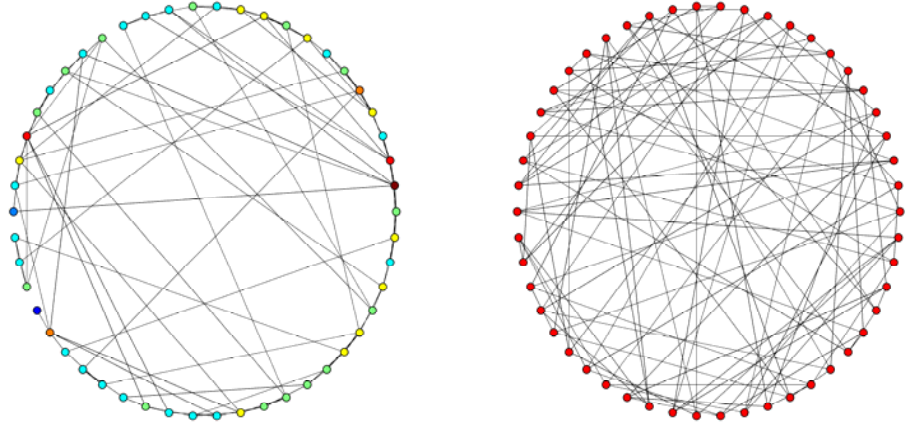

**Figure 2.** The picture shows a Strogatz-Watts network with $N = 50$, $<k> = 4$, and $P = 0.2$ on the left and the resulting optimized network on the right. The colors of nodes in each picture are proportional to their degree, i.e. in each of the pictures; the nodes with same color have the same degree. For the original network in the left $\lambda_N/\lambda_2 = 15.501$, where for the optimized network in the right $\lambda_N/\lambda_2 = 4.948$.

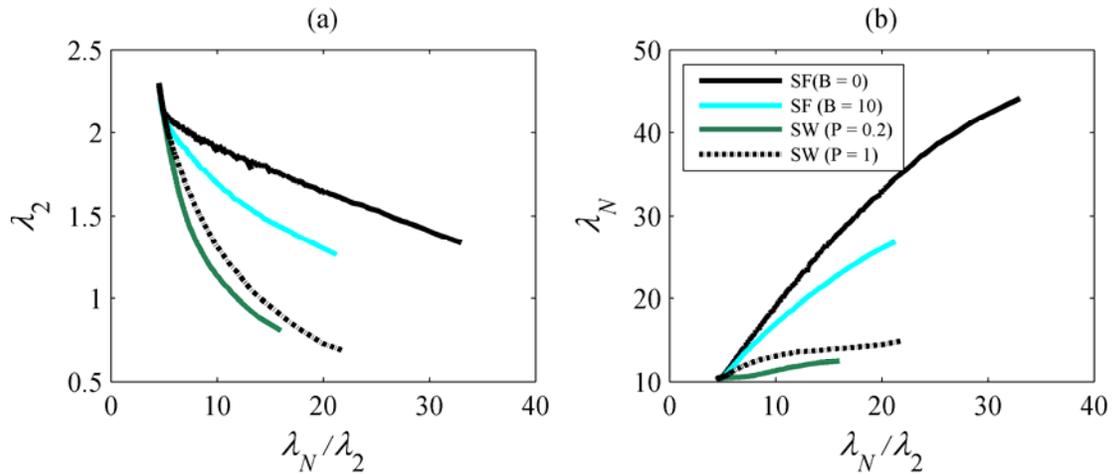

**Figure 3.** Behavior of a) $\lambda_2$ and b) $\lambda_N$ as a function of the eigenratio $\lambda_N/\lambda_2$ during the optimization process for different networks with $N = 200$ and $<k> = 6$.

Figure 4 shows evolution of different parameters of the networks during the optimization process. It can be seen that the networks belong to a class of homogenous networks where the optimized network becomes homogenous in the sense that parameters such as degree, node and edge betweenness centrality vary little over the network. In other words, networks with optimized synchronizability belong to a class of random homogenous network with almost zero variance of degree, node and edge betweenness centrality. Additionally, the maximum node and edge betweenness centrality is distinctly lower than for other types of homogenous networks such as ring or lattice.

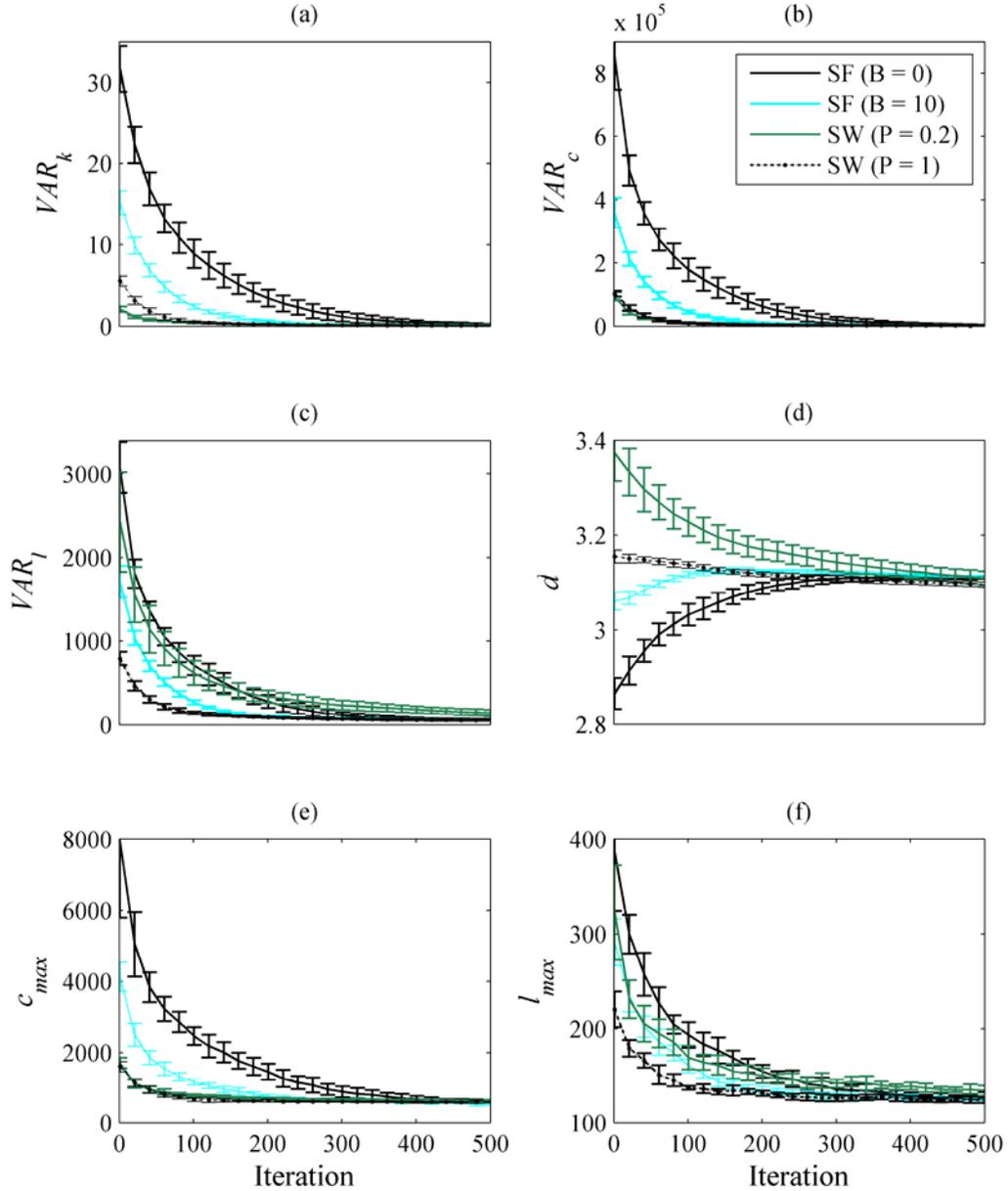

**Figure 4. Profile of a) variance of node degree (VAR$_k$), b) variance of node betweenness centrality (VAR$_c$), c) variance of edge betweenness centrality (VAR$_l$), d) average distance (*d*), e) maximum node betweenness centrality ($c_{max}$), and f) maximum edge betweenness centrality ($l_{max}$), as a function of iteration steps for different networks with $N$ = 200 and $<k>$ = 6. The optimization target is to minimize the eigenratio $\lambda_N/\lambda_2$. Graphs show the averages over 10 realizations with the corresponding errorbars for the standard deviation.**

We investigated also the efficiency of the rewirings. As mentioned, a drawback of the method proposed by Donetti et al. [24] is that the rewirings are done in a blind way that makes the convergence of the optimization algorithm very slow. This problem is overcome in our proposed rewiring algorithm. Figure 5 shows the percentage of the rewiring candidates that makes the target function decrease, and hence are accepted. The results show the average percentage over intervals of iteration steps for different networks with $N = 200$ and $<k> = 6$. It shows that in the first 50 steps, a vast majority of rewirings are accepted, which indeed indicates that the algorithm decreases the eigenratio dramatically in the first few steps. Essentially, as the iterations proceed and a minimum of the target function is approached, less and less candidate rewirings are accepted. These results reveal the effectiveness of the proposed rewirings in optimizing the synchronizability.

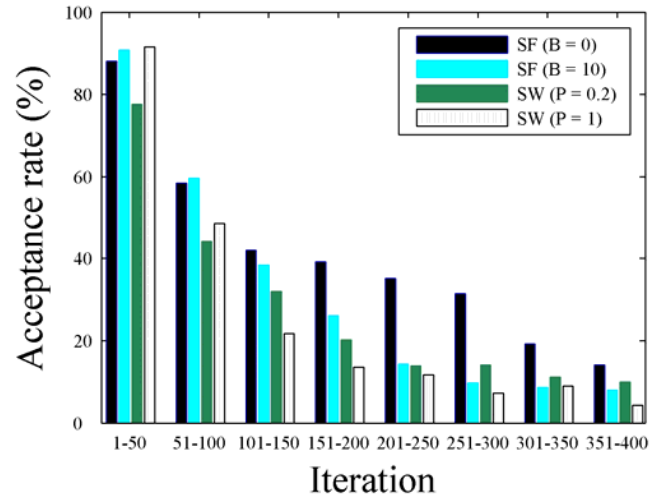

**Figure 5. Percentage of the rewirings which are accepted. The acceptance rates are obtained by averaging over intervals of iteration steps. The networks are with $N = 200$ and $<k> = 6$, and data refer to averaging over 10 realizations.**

The proposed rewiring algorithm is rather fast and can be applied to optimize large networks. Figure 6a (6b) shows the performance of the algorithm for different networks with $<k> = 6$ and $N = 500$ ($N = 1000$). As mentioned, the rewiring algorithm is able to optimize the network to reach reasonable synchronizability in only $2N$ steps. Note that if random rewiring strategy [24] is used, the optimization process, if it works, takes a very long time for large networks. It is worth mentioning that the proposed method is not sensitive to the network size and average degree, and the general behavior of the algorithm remains the same regardless of these properties.

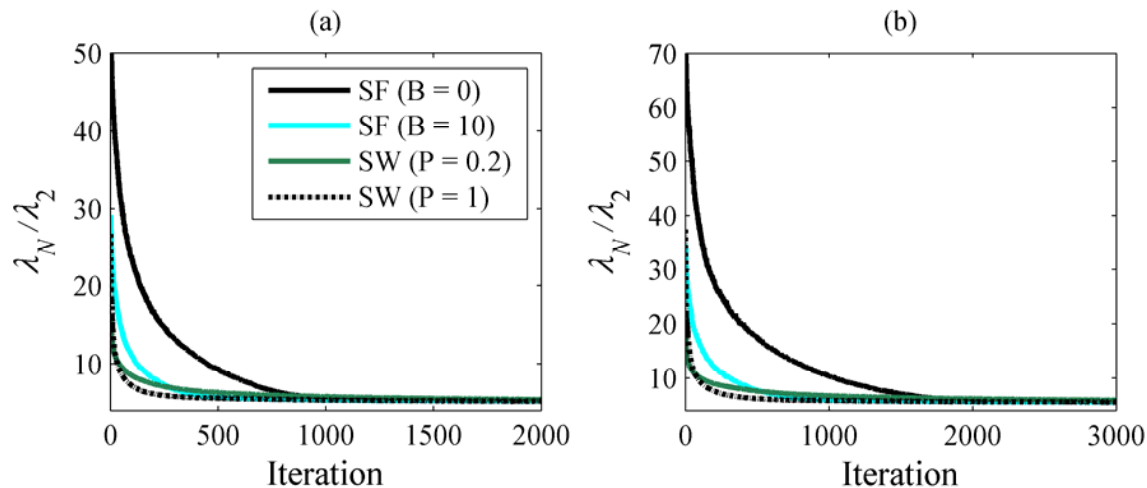

**Figure 6.** The eigenratio $\lambda_N/\lambda_2$ as the iteration steps for different networks with $<k> = 6$ and a) $N = 500$, b) $N = 1000$. Data refer to averages over 10 realizations.

We investigated also the correlation of some network properties such as node degree (number of links connecting to a node), node betweenness centrality (number of shortest paths making use of a node) and edge betweenness centrality (number of shortest paths passing through an edge), with the eigenratio during the optimization period. Table 1 shows the average correlations for networks with $N = 200$ and $<k> = 6$. For scale-free networks the parameter $B$ varied from 0 to 10 and for Strogatz-Watts networks $P$ varied in the range [0.1-1]. One can see that variance of node degree ($VAR_k$), variance of node betweenness centrality ($VAR_c$), variance of edge betweenness centrality ($VAR_l$), maximum degree ($k_{max}$), maximum node betweenness centrality ($c_{max}$), and maximum edge betweenness centrality ($l_{max}$), are highly correlated with $\lambda_N/\lambda_2$. In other words, by minimizing the eigenratio $\lambda_N/\lambda_2$, some network properties are also minimized. This fact lets us employ other optimization targets. For example, we investigated two other optimization goals: minimizing $1/\lambda_2$ and minimizing $c_{max}(VAR_c+1)$ and the performance is shown in Figure 7. As it can be seen, the general behavior is the same as the case with $\lambda_N/\lambda_2$ as optimization target (Figure 1). This indicates that for optimizing the eigenratio $\lambda_N/\lambda_2$, one can take a target function such as $c_{max}(VAR_c+1)$ that avoids computing the eigenvalues at each step. It further indicates that in networks with high synchronizability, different interpretations of synchronizability [5] such as $\lambda_N/\lambda_2$ and $1/\lambda_2$ are almost equivalent.

**Table 1. Pearson correlation coefficients of the $\lambda_N/\lambda_2$ with variance of node degree (VAR$_k$), variance of node betweenness centrality (VAR$_c$), variance of edge betweenness centrality (VAR$_l$), maximum degree ($k_{max}$), maximum node betweenness centrality ($c_{max}$), and maximum edge betweenness centrality ($l_{max}$). The networks are scale-free and Strogatz-Watts networks with $N = 200$ and $<k> = 6$ and the results are averaged over 10 different networks of each type each with 30 realizations.**

|  | $k_{max}$ | VAR$_k$ | $c_{max}$ | VAR$_c$ | $l_{max}$ | VAR$_l$ |
|---|---|---|---|---|---|---|
| Scale-free networks | 0.97 ± 0.005 | 0.99±0.003 | 0.99± 0.003 | 0.99±0.002 | 0.98±0.01 | 0.99±0.002 |
| Strogatz-Watts networks | 0.91±0.051 | 0.95±0.023 | 0.96±0.027 | 0.95±0.014 | 0.93±0.028 | 0.95±0.027 |

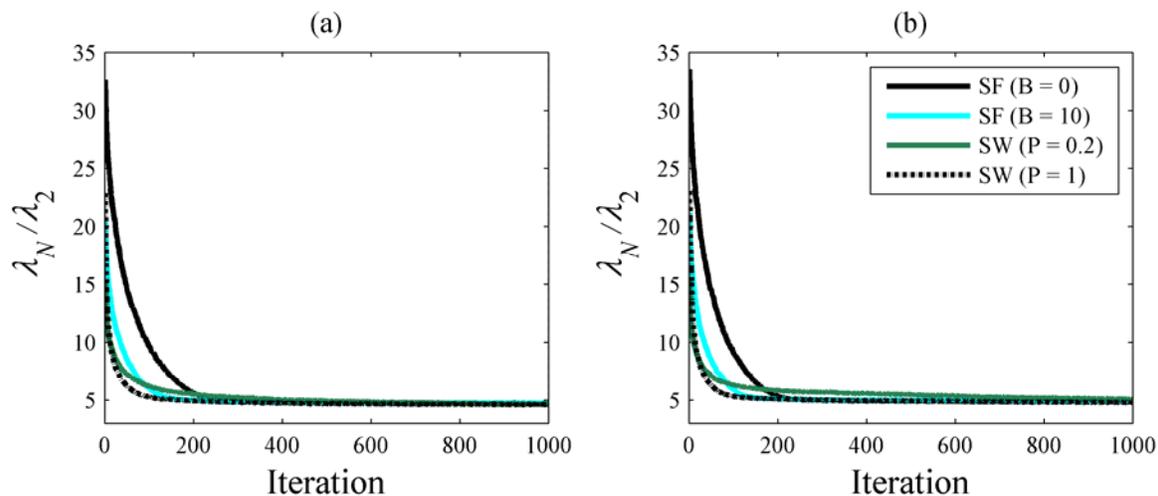

**Figure 7. The eigenratio $\lambda_N/\lambda_2$ as a function of iteration steps for different networks with $N = 200$ and $<k> = 6$. The target functions for optimization are minimization of a) $1/\lambda_2$ and b) $c_{max}(VAR_c+1)$. Graphs show averaging over 10 realizations.**

We compare the performance of the proposed efficient rewiring algorithm with that of the random rewiring proposed in [24]. The results are shown in Figure 8 for scale-free ($B = 0$) and Strogatz-Watts ($P = 0.2$) networks with $N = 200$ and $<k> = 6$. It is seen that the efficient rewiring algorithm could make the eigenratio to reach the steady-state in only 400 iterations, whereas random rewiring algorithm provides the steady-state behavior after 4000 iterations. Thus, the efficient rewiring outperforms in speeding up the optimization task compared to the blind random rewiring strategy. Furthermore, although the optimization is iterated for a long-period (100000 iterations), the network optimized using random rewiring has still larger eigenratio compared to the one optimized using our proposed efficient rewiring. However, it is not excluded that using random rewiring, one might obtain networks with a lower eigenratio (comparable with those of the ones obtained using efficient rewiring), but probably at the price of a large number of repetitions of the optimization process.

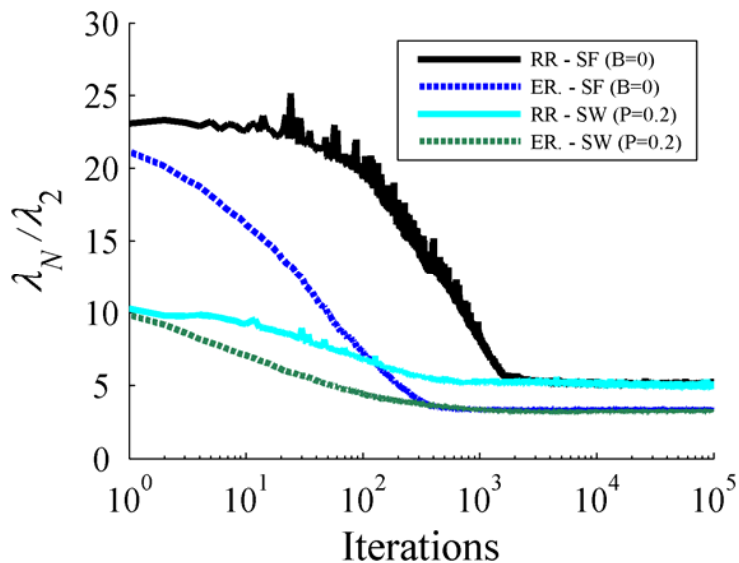

**Figure 8. The eigenratio $\lambda_N/\lambda_2$ as a function of iteration steps for scale-free and Strogatz-Watts network with $N = 200$ and $<k> = 6$. The optimization target is to minimize the eigenratio $\lambda_N/\lambda_2$ and two rewiring algorithms, i.e. random rewiring (RR) and efficient rewiring (ER), are used. Data is averaged over 10 realizations.**

Let us say some words on the complexity of the algorithm. Since the method is based on the calculation of the eigenvalues and eigenvectors of the Laplacian of the connection network, its computational complexity is $O(N^3)$. However, if the network is sparse, which is the case for many applications, there are some methods that calculate only some smallest and largest eigenvalues and the corresponding eigenvectors [38]. A possible extension of the algorithm is to construct it based on only information obtained from node and edge betweenness centrality distributions, which has the computational complexity of $O(NE)$ [39], where $E$ is the number of edges of the network. For sparse networks, the complexity becomes approximately $O(N^2)$ and makes it suitable for applying to very large networks. By choosing the rewiring criteria as $E_{connect,ij} = C_{ij}$ ($C_{ij}$ is the sum of the betweenness centrality of the nodes belonging to the shortest path of the $i$-th and $j$-th node) and $E_{cut,ij} = (k_i + k_j)$, and the target function as $c_{max}(\text{VAR}_c+1)$, the obtained eigenratio $\lambda_N/\lambda_2$ of the optimized network is very close to the one obtained through the eigenvalue based approach (not reported here).

## Conclusions and discussion

Choosing the eigenratio $\lambda_N/\lambda_2$ as synchronizability measure, we proposed a rewiring algorithm for optimizing the synchronizability of dynamical networks. The algorithm employs rules based on the properties of the network for rewiring, i.e. a rule based on the eigenvector corresponding to the second smallest eigenvalue of the Laplacian matrix of the network for adding edges and a rule based on node degree for disconnecting the edges. We utilized a modified version of the simulated annealing approach to perform the optimization task. Since the rewirings are performed in an intelligent fashion, it is much faster compared to the other methods with random rewiring strategy [24, 25]. Our experience showed that roughly $2N$ of rewiring steps is enough to obtain a network with near-optimal synchronizability. We also showed that the algorithm is not sensitive to the specific target function, i.e. $\lambda_N/\lambda_2$, and some other quantities such as $1/\lambda_2$ can also be used equivalently as the target function to optimize the synchronizability. Indeed, in the optimized networks, different interpretations of synchronizability such as "the smaller the $\lambda_N/\lambda_2$ the more synchronizable the network" and "the larger the $\lambda_2$ the more synchronizable the network", coincide.

Starting from any initial network and by employing the proposed rewiring algorithm we end up with a class of homogenous random networks. The optimized networks are homogenous in their degree, node and edge betweenness centrality distribution. Their maximum node and edge betweenness centrality is low and the shortest loop (girth) is large. Also, the possible transitivity and modular structure of the networks vanished during the optimization process.

Since the computation of eigenvalues and the corresponding eigenvectors are rather expensive, one can construct the algorithm based on the node betweenness centrality that is simpler to compute, especially for sparse networks. The betweenness centrality based rewiring algorithms obtains close results compared to the original eigenvalue based approach.

As a consequence of maximizing $\lambda_2$ and perfect homogeneity, the optimized networks belong to a family of networks called Ramanujan networks. A *k*-regular network with the property $\lambda_2 \geq <k> - 2\sqrt{<k>-1}$ is called Ramanujan network [40]. Therefore, our proposed rewiring algorithm can be applied to construct Ramanujan networks of any size and average degree.

## Acknowledgment

This work has been supported by Swiss National Science Foundation through grants No 200020-117975/1 and 200021-112081/1.

## References

1. A. Pikovsky, M. Rosenblum, and J. Kurths, *Synchronization: a universal concept in nonlinear sciences* (Cambridge University Press, 2003).
2. R. Brown and L. Kocarev, Chaos **10**, 344 (2000).
3. V. N. Belykh, I. V. Belykh, and M. Hasler, Physica D **195**, 159 (2004).
4. L. M. Pecora and T. L. Carroll, Physical Review Letters **80**, 2109 (1998).
5. M. Jalili, A. Ajdari Rad, and M. Hasler, International Journal of Circuit Theory and Applications **35**, 611 (2007).
6. G. Korniss, M. A. Novotny, H. Guclu, et al., Science **299**, 677 (2003).
7. S. Barbarossa and G. Scutari, IEEE Transactions on Signal Processing **55**, 3456 (2007).
8. M. E. J. Newman, SIAM Review **45**, 167 (2003).
9. M. Newman, A.-L. Barabasi, and D. J. Watts, *The structure and dynamics of networks* (Princeton University Press, 2006).

10  S. Boccaletti, V. Latora, Y. Moreno, et al., Physics Reports **424**, 175 (2006).
11  R. Albert and A.-L. Barabasi, Reviews of Modern Physics **74**, 47 (2002).
12  S. H. Strogatz, Nature **410**, 268 (2001).
13  D. J. Watts and S. H. Strogatz, Nature **393**, 440 (1998).
14  A.-L. Barabasi and R. Albert, Science **286**, 5009 (1999).
15  G. V. Osipov, J. Kurths, and C. Zhou, *Synchronization in Oscillatory Networks* (Springer, 2007).
16  L. M. Pecora and T. L. Carroll, Physical Review Letters **64**, 821 (1990).
17  S. Boccaletti, J. Kurths, G. Osipov, et al., Physics Reports **366**, 1 (2002).
18  R. Olfati-Saber, J. A. Fax, and R. M. Murray, Proceedings of the IEEE **95**, 215 (2007).
19  M. Chavez, D.-U. Hwang, A. Amann, et al., Physical Review Letters **94**, 218701 (2005).
20  M. Jalili, A. Ajdari Rad, and M. Hasler, Submitted (2008).
21  A. E. Motter, C. S. Zhou, and J. Kurths, Europhysics Letters **69**, 334 (2005).
22  X. Wang, Y.-C. Lai, and C. H. Lai, Physical Review E **75**, 056205 (2007).
23  M. Jalili, A. Ajdari Rad, and M. Hasler, in *IEEE International Symposium on Circuits and Systems*, Seattle, Washington, USA 2008).
24  L. Donetti, P. I. Hurtado, and M. A. Munoz, Physical Review Letters **95** 188701 (2005).
25  L. Donetti, F. Neri, and M. A. Munoz, Journal of Statistical Mechanics: Theory and Experiment (2006).
26  A. E. Motter, C. Zhou, and J. Kurths, Physical Review E **71** 016116 (2005).
27  T. Nishikawa, A. E. Motter, Y.-C. Lai, et al., Physical Review Letters **91**, 014101 (2003).
28  H. Hong, M. Y. Choi, and B. J. Kim, Physical Review E **65**, 026139 (2002).
29  X. F. Wang and G. Chen, IEEE Transactions on Circuits and Systems—I: Regular Papers **49**, 54 (2002).
30  X. F. Wang and G. Chen, International Journal of Bifurcation and Chaos **12**, 187 (2002).
31  M. Barahona and L. M. Pecora, Physical Review Letters **89**, 054101 (2002).
32  H. Hong, B. J. Kim, M. Y. Choi, et al., Physical Review E **69**, 067105 (2004).
33  L. Donetti, P. I. Hurtado, and M. A. Munoz, Journal of Physics A: Mathematical and Theoretical (2008).
34  B. Wang, H. Tang, C. Guo, et al., Physica A: Statistical Mechanics and its Applications **368**, 607 (2006).
35  S. Fallat and S. Kirkland, Electronic Journal of Linear Algebra **3**, 48 (1998).
36  A. Ghosh and S. Boyd, in *IEEE Conference on Decision and Control*, 2006), p. 6605.
37  W. N. Anderson and T. D. Morley, Linear and Multilinear Algebra **18**, 141 (1985).
38  J. W. Demmel, *Applied numerical linear algebra* (SIAM, 1997).
39  U. Brandes, Journal of Mathematical Sociology **25**, 163 (2001).
40  G. Davidoff, P. Sarnak, and A. Valette, *Elementary number theory, group theory and Ramanujan graphs* (Cambridge University Press, 2003).